\documentclass[preprint,tightenlines,floatfix,
nofootinbib,eqsecnum,superscriptaddress]{revtex4-1}

\usepackage{amsfonts,amstext,mathrsfs}

\usepackage{commath}
\usepackage{braket}
\usepackage{subfigure}
\usepackage{multirow}
\usepackage{tabularx}
\usepackage{pstricks}
\usepackage[section]{placeins}
\usepackage{booktabs}
\usepackage{array}

\usepackage{mathpazo}

\usepackage{graphicx}
\usepackage{url}
\usepackage{amsbsy,amssymb} 
\usepackage{amsmath}
\usepackage{hyperref}
\usepackage{color}

\usepackage{feynmf}
\newcommand{\bp}{\mbox{\boldmath $p$}}

\newcommand{\bpaa}{\mbox{\boldmath $p_{1}$}}
\newcommand{\bpbb}{\mbox{\boldmath $p_{2}$}}
\newcommand{\bpbc}{\mbox{\boldmath $p_{3}$}}
\newcommand{\bpab}{\mbox{\boldmath $p_{12}$}}

\newcommand{\bhp}{\mbox{\boldmath $\hat{p}$}}

\newcommand{\bsigma}{\mbox{\boldmath $\sigma$}}

\renewcommand\slash[1]{\not \! #1}

\usepackage[normalem]{ulem}  

\usepackage{fancyvrb}

\newcommand{\codeMark}[1]{{\protect\Verb+#1+}} 
\newcommand{\commandMark}[1]{{\protect\Verb+#1+}} 
\newcommand{\classMark}[1]{{\protect\Verb+#1+}} 

\newcommand{\dirFileMark}[1]{{\it #1}} 

\begin{document}

\title{\textsc{Decay}: A Monte Carlo library for the decay of a particle with ROOT compatibility}
%
%
 
\vspace{0.6cm}

\author{R.A. Kycia}
\email{kycia.radoslaw@gmail.com}
\affiliation{Masaryk University, Department of Mathematics and Statistics, Kotl\'{a}\v{r}sk\'{a} 267/2, 611 37 Brno, The Czech Republic}
\affiliation{Cracow University of Technology, Faculty of Materials Science and Physics, Warszawska 24, PL-31155 Krak\'ow, Poland}

\author{P. Lebiedowicz}
\email{Piotr.Lebiedowicz@ifj.edu.pl}
\affiliation{Institute of Nuclear Physics Polish Academy of Sciences, Radzikowskiego 152, PL-31342 Krak{\'o}w, Poland}

\author{A. Szczurek
\footnote{Also at \textit{College of Natural Sciences, 
Institute of Physics, University of Rzesz{\'o}w, 
ul. Pigonia 1, PL-35310 Rzesz{\'o}w, Poland}.}}
\email{Antoni.Szczurek@ifj.edu.pl}
\affiliation{Institute of Nuclear Physics Polish Academy of Sciences, Radzikowskiego 152, PL-31342 Krak{\'o}w, Poland}

\begin{abstract}
\noindent
Recently, there is a need for a general-purpose event generator of decays of an elementary particle 
or a hadron to a state of higher multiplicity ($N > 2$) that is simple to use and universal. 
We present the structure of such a library to produce generators that generate kinematics of decay processes 
and can be used to integrate any matrix element squared over phase space of this decay. 
Some test examples are presented, and results are compared with results known from the literature.
As one of examples we consider the Standard Model Higgs boson decay into four leptons.
The generators discussed here are compatible with the ROOT interface.
\end{abstract}

\maketitle

\section{Introduction}
In modern high energy physics or hadronic physics, 
the prime role is played by complex multiprocess Monte Carlo generators. 
One example of such a versatile tool 
is \textsc{Pythia} \cite{Sjostrand:2014zea,Sjostrand:2006za}.

However, apart from beam colliding experiments, there are various
experiments (e.g., \cite{Aaltonen:2015uva,Adam:2020sap,
Sirunyan:2020cmr,Sikora_PhD_thesis})
that measure decaying unstable particle products. 
In many cases, there are specific generators for such processes, 
with specific matrix elements and kinematics tuned 
for the particular decay.

However, in some theoretical investigation, there is a need for a simple
yet versatile generator for prototyping new theories or to understand
mechanism of decays of mesons or baryons.
Good examples are exotic decays of conventional mesons, 
decays of glueballs or tetraquarks.
One typical choice is the \textsc{GENBOD} code \cite{James:1968} which uses 
the Raubold and Lynch algorithm. It is delivered in ROOT package
\cite{ROOT} as \classMark{TGenPhaseSpace} class. There are some 
inconveniences in the use of this generator. The events are weighted,
and it requires some work to integrate some expression over a phase
space and generate events. Moreover, some additional tools must be added
to provide adaptive Monte Carlo integration/simulation.

These inconveniences were present during our theoretical investigations
of the phenomenology of decaying processes 
\cite{Lebiedowicz:2016zka,Lebiedowicz:2019jru},
where the $1 \to 4$ decay processes were simulated 
from the $1 \to 2$ process with a corresponding spectral functions.
This motivated us to constuct a new, 
more powerful Monte Carlo generators for decays. 
They should be compatible with ROOT software and, due to adaptivity, can
handle with integrands that have 'small' support in 
the Lorentz Invariant Phase Space (LIPS). These tools are based on 
some elements from our previous exclusive MC generator 
\textsc{GenEx} \cite{Kycia:2014hea,Kycia:2017ota}. The essential
requirement in designing these tools is that the interface should be 
fully compatible with ROOT generator \classMark{TGenPhaseSpace} and 
use effectively other ROOT components.

One of the application of these tools will be to use them
in central exclusive diffractive production 
of $\pi^{+}\pi^{-}\pi^{+}\pi^{-}$ 
\cite{Lebiedowicz:2016zka,Kycia:2017iij} and
$K^{+} K^{-} K^{+} K^{-}$ \cite{Lebiedowicz:2019jru}
in proton-proton collisions that proceeds via resonances.
These processes are under recent experimental studies 
at RHIC and LHC \cite{Adam:2020sap,Sirunyan:2020cmr,Sikora_PhD_thesis}.

Another application is their use in search of hypothetical resonances of new physics 
and study their properties, 
e.g., in the four-lepton channel \cite{Aaboud:2017rel,Aad:2020fpj}.
The $H \to ZZ \to 4 \ell$ channel, as compared to other final states,
has the advantage of being experimentally clean and,
for this reason, is considered 
the ``golden'' channel to explore the possible
existence of a heavy Higgs resonance.
As an example, we consider the decay $H \to ZZ \to 4 \ell$
of spin~0 Higgs boson with mass $M_{H} = 125.1$~GeV.

The paper is organized as follows: In the next section, 
implementation details and usage of the library will be presented. 
The package also contains three generators for decays with two, 
three, and four particles in the final states. 
The physical process of these kinds for testing purposes 
were described in the second part of the paper. 
These are the application of $4$-Fermi theory to $\mu$ decay
and the Standard Model Higgs boson decay to four leptons.

\section{Implementation details}

\subsection{General requirements and running \textsc{Decay}}
The requirements for running the software are:

\begin{itemize}
 \item {ROOT library \cite{ROOT} - 
 the set of generators integrates with this library and uses many of its components.}
 \item {$C/C++$ compiler - 
 the default compiler is the one with GNU GCC collection \cite{GCC}.}
 \item {GNU MAKE \cite{GNUMAKE} - 
 examples are provided with fully functional \dirFileMark{Makefile}s.}
 \item {Doxygen \cite{Doxygen} - 
 for generating documentation from the code.}
\end{itemize}

The examples are supplied with \dirFileMark{Makefile} with the following commands defined:
\begin{itemize}
 \item {\commandMark{make run} - compiles and runs the example.}
 \item {\commandMark{make clean} - cleans executables.}
 \item {\commandMark{make cleanest} - cleans executables and results of simulation.}
 \item {\commandMark{make Generate-doc} - generates documentation 
 in pdf and LaTeX file using Doxygen.}
\end{itemize}

\subsection{How to install \textsc{Decay}}
The library and examples can be downloaded from the repository \cite{GitHub:2020}.
The repository is split into two directories:
\begin{itemize}
 \item {\dirFileMark{Library} - contains header files and implementations of library. 
 These files can be placed inside the user directory or placed in common place for compiler.}
 \item {\dirFileMark{Examples} - contains three examples described below in details. 
 They illustrate the use of files from library in a real-life applications:}
	\begin{itemize}
 	\item {\dirFileMark{2DDecay} - decay of a central blob into two particles; 
 	see Section~\ref{Section_2Particles} for details.}
 	\item {\dirFileMark{4Fermi} - $\mu \to \nu_{\mu} e \bar{\nu}_{e}$ decay 
 	in 4-Fermi phenomenological theory \cite{Schwartz:2013pla}; 
 	see Section~\ref{Section_3Particles} for details.}
 	\item {\dirFileMark{H4} - contains two models of $H$ decay to four leptons described below; 
 	see Section~\ref{Section_4Particles} for details.}
	\end{itemize}
\end{itemize}

In order to use the content of \dirFileMark{Library} directory you have to write your own C++ \dirFileMark{main.cxx} file that includes specific files from the directory. A detailed description is provided below. 

It is however advised to experiment with specific examples beforehand. They are stand-alone programs that are using \textsc{Decay} library from \dirFileMark{Library} directory. The examples can be used as a starting point for writing the generator for a new process economically.

Each example is supplied with \dirFileMark{Makefile} that simplifies process of compiling and running. 
The commands to run the examples are as follows:
\begin{itemize}
 \item {\commandMark{make run} - compiles and runs the example.}
 \item {\commandMark{make clean} - cleans executables.}
 \item {\commandMark{make cleanest} - cleans executables and results of simulation.}
 \item {\commandMark{make Generate-doc} - generates documentation 
 in pdf and LaTeX file using Doxygen.}
\end{itemize}

\subsection{Input}

The examples require no specific input, and all specific parameters and matrix elements of physical models are present in corresponding \dirFileMark{main.cxx} files.

\subsection{Output}

Each generator in the discussed here examples generates four-vectors of
particles from the decay. 
They are saved in the file \dirFileMark{events.txt} on disk 
and some control plots of rapidity and invariant mass in eps/pdf files,
and, in addition, the file \dirFileMark{histograms.root} 
to create a histograms in ROOT.
These are standard ways how to deal with events 
using straightforward approach known from the ROOT package. 
All the logic is contained in \dirFileMark{main.cxx} file.

\subsection{Two modes of using examples of \textsc{Decay}}
\label{subsec_two_modes_of_using}

The presented here examples of \textsc{Decay} package can be used in two distinct ways:
\begin{itemize}
 \item[(A)] First events are generated according to phase-space distributions.
Then the events are weighted by the square of the matrix element
${\cal M}_{1 \to n}$ for the $1 \to n$ decay 
prepared by the user
outside of the generator in a separate program.
This option may be easier for the user but difficult for
controlling accuracy.
 \item[(B)] The matrix element squared for the decay is inserted directly into
the generator itself. This option may be more optimal as far as
efficiency and accuracy are considered.
\end{itemize}

\subsection{Structure of classes}
In this section, the structure of the system of generators will be
presented. This description is related to 
the files in \dirFileMark{Library} directory and is aimed at users who
want to write the generator based on this library from scratch.

The library uses the modified \codeMark{TDecay} class from previous generator 
for exclusive processes \textsc{GenEx} 
\cite{Kycia:2014hea,Kycia:2017ota} 
that is based on the original \textsc{GENBOD} algorithm \cite{James:1968}. 
The interface is based on \codeMark{TGenPhaseSpace} class from ROOT 
for compatibility. 
The difference is that
\begin{itemize}
 \item {It requires (pseudo)random numbers to generate an event. 
This is needed to connect this generator with an external system 
that feeds this numbers, e.g., adaptive sampling as described below.}
 \item {It returns the volume of the Lorentz Invariant Phase Space
     (LIPS) \cite{James:1968,Hagedorn:1964,Pilkuhn:1967,Schwartz:2013pla}
\begin{equation}
d^{(n)}{\rm LIPS} = d{\rm LIPS} \left(P \rightarrow \sum_{i=1}^{n}p_{i}\right) = (2\pi)^4 \delta^{(4)}\left(P - \sum_{i=1}^{n}p_{i}\right) \prod_{i=1}^{n}\frac{d^{3}p_{i}}{(2\pi)^{3}2E_{i}}\,,
\label{Eq.LIPS}
\end{equation}
where $P$ is the 4-momentum of decaying particle, 
$p_{i}$ are 4-momenta of the decay products and $E_{i}$ their energies. }
\end{itemize}

This class follows the general interface described 
in the following subsection.

\subsection{Interface description}
\begin{figure}[!ht]
\centering
 \includegraphics[width=.5\textwidth]{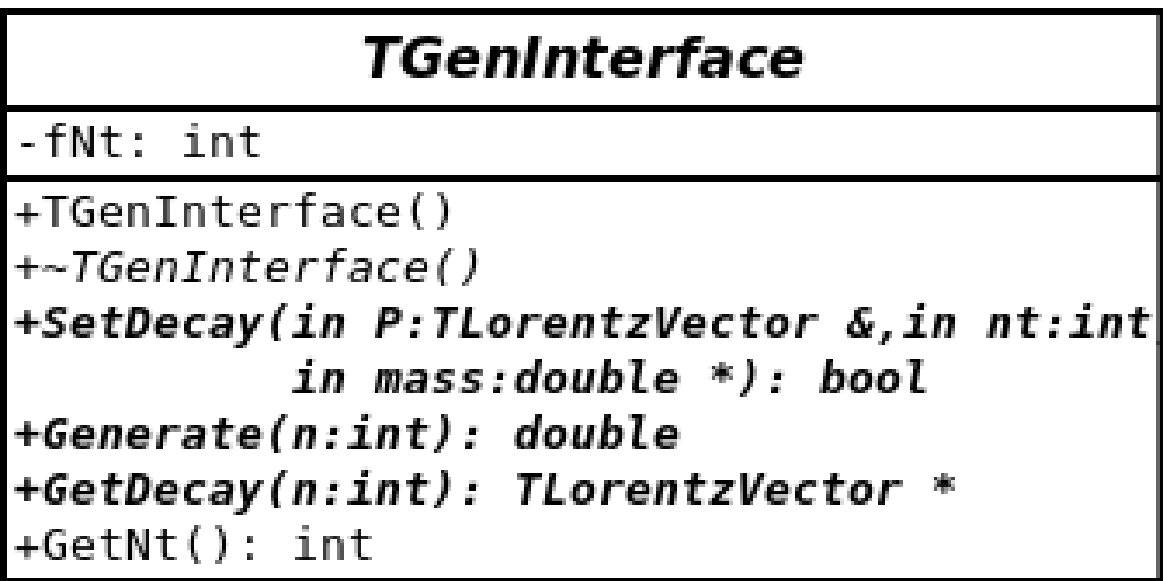}
 \caption{Interface for generators.}
 \label{Fig.TGenInterface}
\end{figure}

In this subsection, we describe the interface of the discussed decay 
generators. 
The interface is derived from the standard ROOT generator
\codeMark{TGenPhaseSpace} to make it compatible with this package.
The interface is enclosed in the abstract class 
\codeMark{TGenInterface} that is presented in Fig.~\ref{Fig.TGenInterface}.
The description of methods is as follows:
\begin{itemize}
 \item {\codeMark{SetDecay(TLorentzVector \&P, Int\_t nt, const Double\_t *mass)} 
 - set decay configuration. The initial 4-momentum of a particle is $P$, 
 $nt$ is the number of final state particles, and $mass$ is an array of masses of final particles. This is the first method that should be called before generating an event.
 It returns \codeMark{true} when configuration 
 is allowed and \codeMark{false} otherwise.}
 \item {\codeMark{Generate(void)} generates an event and returns 
 the weight of this event. It should be called after the configuration is set.}
 \item {\codeMark{GetDecay(Int\_t n)} returns the $n$-th product of the decay 
 as a 4-vector.}
\end{itemize}

When using object that realizes above interface, the sequence of calls should as follows:
\begin{enumerate}
 \item {\codeMark{SetDecay(...)}}
 \item {In the loop:}
 \begin{enumerate}
  \item {\codeMark{Generate(...)}}
  \item {\codeMark{GetDecay(...)}}
 \end{enumerate}
\end{enumerate}
This mimic the sequence of calls from well known \classMark{TGenPhaseSpace} class of ROOT for compatibility.

This interface is implemented in the following generators.

\subsection{TDecay}
The \codeMark{TDeacy} class contains specialization of the \codeMark{TGenDecay}
specific methods with specialization. 
The method \codeMark{Generate (std::queue< double > \&rnd)} 
takes the queue \codeMark{rnd} of $3n-4$ random numbers 
needed for generation of events.
The weight returned by this method is precisely (\ref{Eq.LIPS}).

This class is the backbone of more advanced generators described below.

\subsection{TGenDecay}
\begin{figure}[!ht]
\centering
 \includegraphics[width=.5\textwidth]{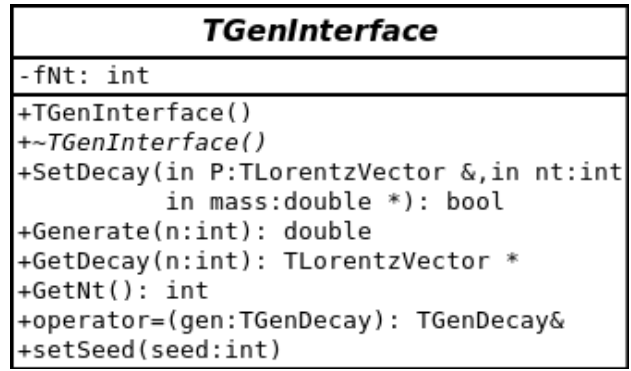}
 \caption{The diagram of class \classMark{TGenDecay}.}
 \label{Fig.TGenDecay}
\end{figure}

Wrapping of \codeMark{TDeacy} with the uniform random number generator 
is \codeMark{TGenDecay}.
Its class diagram is presented in Fig.~\ref{Fig.TGenDecay}.
Apart of the methods from the \classMark{TGenInterface}, 
there is an additional method \codeMark{setSeed(...)}, 
which sets the seed for internal pseudorandom number generator.

\begin{figure}[!ht]
\centering
 \includegraphics[width=.5\textwidth]{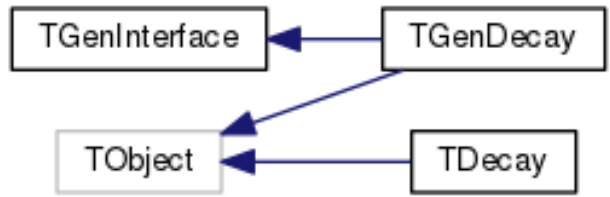}
 \caption{Inheritance diagram for \classMark{TGenDecay}.}
 \label{Fig.TGenDecayInheritance}
\end{figure}

The inheritance diagram is presented in Fig.~\ref{Fig.TGenDecayInheritance}.
The inheritance from \classMark{TObject} from the ROOT class is 
for compatibility.

The typical use of this generator is as follows:
\begin{itemize}
 \item {\codeMark{TGenDecay generator;} - makes an instance of the generator;}
 \item {\codeMark{generator.SetDecay(P, nt, mass);} - sets the decay configuration;}
 \item {In the loop over events ($N$-times):}
	\begin{itemize}
  \item {\codeMark{w =  generator.Generate();} - 
  makes an event and remembers the volume of the phase space 
  in \codeMark{double w}; 
  Call the LIPS weight for $i$-th event $w_{i}$.}
  \item {\codeMark{pfi = *(generator.GetDecay( i )); } - 
  get the 4-momentum of $i$-th particle; 
  It can be done for all particles needed to compute integrand.}
  \item {Computes the integrand $I_{i}$ for the $i$-th event.}
	\end{itemize}
\end{itemize}
Then by the standard Monte Carlo reasoning \cite{James:1968} 
the integral is
\begin{equation}
\int I(p_{1},\ldots, p_{nt}) \,d{\rm LIPS} \approx <I>  \pm \sigma,
\end{equation}
where
\begin{equation}
<I>=\frac{1}{N}\sum_{i=1}^{N} I_{i}w_{i},
\end{equation}
and the error estimate is
\begin{equation}
\sigma = (<I>^2 - <I^2>)\sqrt{N},
\end{equation}
where 
\begin{equation}
<I^2> = \frac{1}{N}\sum_{i=1}^{N} w_{i} I_{i}^{2}.
\end{equation}

The class \classMark{TGenDecay} is useful for symmetric decay. 
Therefore it can be used for integrating matrix element that is not aspherical since 
then the effectiveness can be low, as it was shown for \classMark{TDecay} 
in \cite{Kycia:2017ota,Kycia:2014hea}. 
If this is the case, then a more advanced decay generator described 
in the next section is more relevant.

\subsection{Adaptive decay with TGenFoamDecay}
There are cases where a simple approach described above is insufficient. 
The first one is when the integrand has the support, which is small in the LIPS region over which we integrate. This includes the integrands, which have strong 'non-spherical' cuts on momenta and energies of decay products. In this case, one can try to integrate over LIPS using the adaptive Monte Carlo procedure. There are two common general-purpose MC integrators VEGAS \cite{Lepage:1977sw,Ohl:1998jn,Lepage:2020tgj} 
and FOAM \cite{Jadach:2002kn}. Merging of the \classMark{TDecay} and FOAM algorithm is made in \classMark{TGenFoamDecay} class.

\begin{figure}[!ht]
\centering
 \includegraphics[width=.5\textwidth]{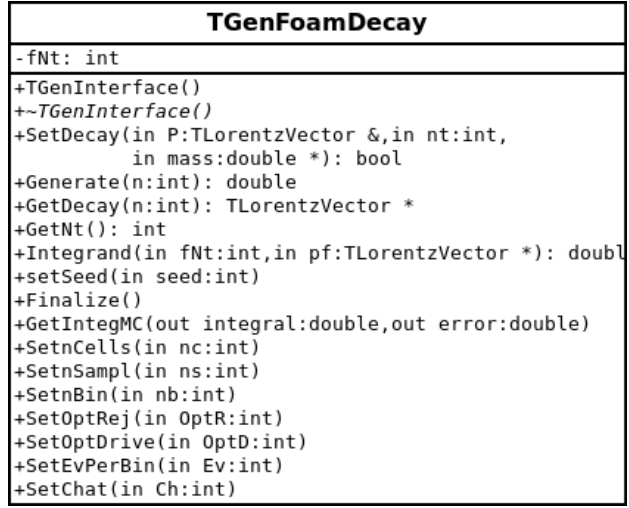}
 \caption{Class diagram for \classMark{TGenFoamDecay}.}
 \label{Fig.TGenFoamDecay}
\end{figure}

The class diagram is presented in Fig.~\ref{Fig.TGenFoamDecay}.
The difference between \classMark{TGenDecay} and \classMark{TGenFoamDecay} 
is due to FOAM content and facilitate adaptive integration. The following methods are new compared to the interface \classMark{TGenInterface}:
\begin{itemize}
 \item {In the \codeMark{SetDecay(...)} method, 
 the initialization of FOAM decay is made.}
 \item {\codeMark{Double\_t Integrand( int fNt, TLorentzVector* pf )} - 
 is a virtual method that fixes integrand that FOAM integrates over LIPS. 
 In the method \codeMark{fNt} is the number of particles, and \codeMark{pf} -
 momenta of the decay products. In this class it returns $1.0$ 
 and can be changed when the user makes a new class that inherits from
 \classMark{TGenFoamDecay}, as it will be explained below.}
 \item {\codeMark{Finalize( void )} - 
 the method that makes finalization of FOAM 
 and prints the integral and its error on the terminal.}
 \item {\codeMark{GetIntegMC(Double\_t \& integral, Double\_t \& error)} - 
 the method that allows to extract the integral 
 and its error from the FOAM integrator inside \classMark{TGenFoamDecay}.}
 \item {The methods \codeMark{Set...} are setting parameters of FOAM integrator
 \cite{Jadach:2002kn}. Default values are good for most standard applications,
 however, if it is not the case these parameters should be set before
 \codeMark{SetDecay(...)} is called. Some practical hints are as follows:}
 \begin{itemize}
  \item {If the matrix element has 'small' support in LIPS 
  and the integral is small, then the \codeMark{nCells} 
  and \codeMark{nSampl} should be increased.}
  \item {By default, \codeMark{TGenFoamDecay} generates the events 
  with weight $1.0$ since \codeMark{OptRej=1}. 
  However, if one prefers weighted events, 
  then one should set \codeMark{OptRej=0}.}
 \end{itemize}
\end{itemize}

Use of \classMark{TGenFoamDecay} in a specific decay with a specific matrix element of reaction is made by deriving a new class from \classMark{TGenFoamDecay} and redefining virtual function
\codeMark{Integrand(..)}. For example, below, the specific \classMark{Generator} class was derived and the \codeMark{Integrand} method was specified.
\begin{verbatim}
class Generator: public TGenFoamDecay 
{
Double_t Integrand( int fNt, TLorentzVector * pf );
};

Double_t Generator::Integrand( int fNt, TLorentzVector * pf )
{
double integrand = 1.0; //Here put your integrand instead of 1.0
return integrand;	
};	
\end{verbatim}

\begin{figure}[!ht]
\centering
 \includegraphics[width=.5\textwidth]{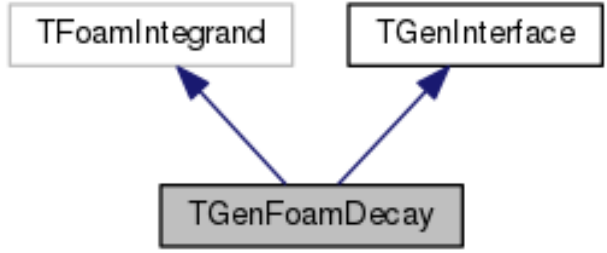}
 \caption{Inheritance diagram for \classMark{TGenFoamDecay}.}
 \label{Fig.TGenFoamDecayInheritance}
\end{figure}

Inheritance diagram is presented in Fig.~\ref{Fig.TGenFoamDecayInheritance}. 
It shows that \classMark{TGenFoamDecay} implements \classMark{TGenInterface}
interface and at the same time is FOAM integrand class \classMark{TFoamIntegrand}.

In the next section, some simple and useful applications and tests of 
the discussed generator will be described.

\section{Tests}
The detailed test of \classMark{TDecay} was presented in \cite{Kycia:2017ota}. 
We present tests of the use of the full generators in examples of decays into 
2, 3, and 4 final state particles. 
Examples that realize these test are contained in \dirFileMark{Examples} 
directory of \cite{GitHub:2020}.

\subsection{Two particle decay}
\label{Section_2Particles}
In the case of a two-particle final state, 
the integral over the whole phase space is known 
in the closed form \cite{Hagedorn:1964, Pilkuhn:1967}, 
namely, in the center of mass of particle $a$ we have

\begin{eqnarray}
R_{2}(a \rightarrow b+c) &=& \int \frac{d^{3}p_{b}d^{3}p_{c}}{(2\pi)^{3}2E_{b}(2\pi)^{3}2E_{c}}(2\pi)^{4}\delta^{(4)}(P_{a}-(p_{b}+p_{c}))
\nonumber \\
&=&\frac{\pi}{2m_{a}^{2}(2\pi)^2}\sqrt{(m_{a}^{2}-(m_{b}+m_{c})^2)(m_{a}^{2}-(m_{b}-m_{c})^2)}\,.
\end{eqnarray}

For a test we considered a decay of a hypothetical decay 
of particles $m_{a}=m_{\mu}$ into $m_{b}=m_{c}=m_{e}$. 
This can serve as a precise test since the formula for the decay is exact in this case.

\subsection{Three particle decay}
\label{Section_3Particles}
In the 4-Fermi theory \cite{Schwartz:2013pla}, 
which is an effective theory of weak interactions, 
the decay $\mu \to e \nu_{\mu} \bar{\nu}_{e}$ 
is given by the simple matrix element \cite{Schwartz:2013pla}
\begin{equation}
|{\cal M}|^{2} = 32\,G_{F}^{2}(m_{\mu}^{2}-2m_{\mu}E_{\bar{\nu}_{e}})m_{\mu}E_{\bar{\nu}_{e}}\,,
\end{equation}
where the Fermi coupling constant $G_{F} = 1.166 379 \times 10^{-5}$~GeV$^{-2}$ makes 
the energy scale of the usefulness of this non-renormalizable 
theory\footnote{There is no problem with non-renormalizability away 
of energy scale $G_{F}^{-2}$, similarly to the usefulness of quantum gravity
perfect predictions away of the Planck energy \cite{Schwartz:2013pla}. 
The theory should be replaced by the theory of weak interactions 
when the scale of $G_{F}^{-2}$ is reached.}, 
$m_{\mu}$ is the mass of muon, 
$E_{\bar{\nu}_{e}}$ 
is the energy of the outgoing electron antineutrino
$\bar{\nu}_{e}$, and 
the masses of neutrinos are negligible 
($m_{\nu_{\mu}} = m_{\bar{\nu}_{e}} = 0$).
The decay width is given by the standard formula
\begin{equation}
\Gamma(\mu \to e \nu_{\mu} \bar{\nu}_{e}) =\frac{1}{2m_{\mu}}\int |{\cal M}|^{2} d^{(3)}{\rm LIPS}\,,
\label{Eq.4FermiGamma}
\end{equation}
which can be calculated analytically to \cite{Schwartz:2013pla}
\begin{equation}
\Gamma(\mu \to e \nu_{\mu} \bar{\nu}_{e}) = \frac{G_{F}^{2}m_{\mu}^{5}}{192\pi^3}\,.
\label{Gamma_2to3}
\end{equation}
This makes a perfect testbed for the generator.  

In case of use \classMark{TGenFoamDecay} class, 
we derive \classMark{Generator} class that redefines 
integrand by (\ref{Eq.4FermiGamma}), namely,
\begin{verbatim}
class Generator: public TGenFoamDecay 
{
Double_t Integrand( int fNt, TLorentzVector * pf );	
};

Double_t Generator::Integrand( int fNt, TLorentzVector * pf )
{
const double G = 1.166e-5;
const double mmu = 0.1057;
double M2 = 32.0*G*G*( mmu*mmu-2.0*mmu*pf[2].E() ) * mmu*pf[2].E();
double corr = 1.0/(2.0*mmu) ;
return M2 * corr;	
};	
\end{verbatim}
where the $\rm pf[2]$ is the \classMark{TLorentzVector} object 
that represents 4-momentum of $\bar{\nu}_{e}$, 
if we define the mass matrix in the following order 
\codeMark{mass = \{$m_{e}$, $m_{\nu_{\mu}}$, $m_{\bar{\nu}_{e}}$\}}.

Table~\ref{Tab.4Fermi_TGenDecay} presents the convergence of 
the Monte Carlo method using \classMark{TGenDecay}. 
It is the typical convergence with $\sim \frac{1}{\sqrt{N}}$ accuracy, 
where $N$ is statistics of events.
\begin{table}
\begin{center}
\begin{tabular}{ |c| c| c| }
\hline
 \# events & $\Gamma_{n}$ [GeV] & $\abs{\Gamma_{n}-\Gamma_{\rm exact}}$ [GeV] \\ \hline 
 $10^2$ & $2.9285\times 10^{-19} $ & $8.465\times 10^{-21}$  \\ \hline
 $10^4$ & $3.00344\times 10^{-19}$ & $9.71 \times 10^{-22}$  \\ \hline
 $10^6$ & $3.01232\times 10^{-19}$ &  $8.3 \times 10^{-23} $ \\ \hline
\end{tabular}
\end{center}
\caption{The value of $\Gamma_{n}$ (\ref{Eq.4FermiGamma})
obtained by \classMark{TGenDecay} integration.
The exact value (\ref{Gamma_2to3}) 
is $\Gamma_{\rm exact} = 3.01315 \times 10^{-19}$~GeV.}
 \label{Tab.4Fermi_TGenDecay}
\end{table}

\subsection{Four particle decay}
\label{Section_4Particles}
In this subsection, we describe a test of the decay 
of the Standard Model Higgs boson into four leptons $H \rightarrow 4 \ell$, 
which is of interest in precision tests of Higgs boson at the LHC 
and in future colliders.

The decay rate for $H \rightarrow 4 \ell$ can be written as
\begin{eqnarray}
\Gamma(H \to 4 \ell) =
\frac{1}{2 M_{H}} \frac{1}{F}
\int {\abs{{\cal M}_{H \to 4 \ell}}^{2}} d^{(4)}{\rm LIPS} \; ,
\label{Gamma_H4L}
\end{eqnarray}
where $1/F$ is the identity factor of particles in the final state
($F = 1$ for $e^{+}e^{-}\mu^{+}\mu^{-}$ and
$F = 4$ for $\mu^{+}\mu^{-}\mu^{+}\mu^{-}$).

\subsubsection{The matrix element for $H \to 4 \ell$ decay}
\label{subsec_Hto4L_explicite}

Here we consider the decay amplitudes of the Higgs boson 
into four leptons in two processes.

For the decay
\begin{eqnarray}
H(P) \to e^{+}(p_{1}) \,e^{-}(p_{2})\,\mu^{+}(p_{3})\,\mu^{-}(p_{4})
\label{H_2e2mu}
\end{eqnarray}
via the intermediate $Z$ bosons, the amplitude reads
\begin{eqnarray}
&&{\cal M}_{H \to e^{+}e^{-}\mu^{+}\mu^{-}} = e^{3}\,g_{HZZ}
\nonumber \\
&&\qquad \times \frac{1}{(p_1 + p_2)^2 - M_Z^2 + i M_Z \Gamma_Z}
\,\bar{u}_{s_{2}}(p_{2})
[\gamma^{\mu}(g_{Z\ell\ell}^{+} \omega_{+} + g_{Z\ell\ell}^{-} \omega_{-})] 
v_{s_{1}}(p_{1})  \nonumber \\
&&\qquad \times  \frac{1}{(p_3 + p_4)^2 - M_Z^2 + i M_Z \Gamma_Z}
\,\bar{u}_{s_{4}}(p_{4})
[\gamma_{\mu}(g_{Z\ell\ell}^{+} \omega_{+} + g_{Z\ell\ell}^{-} \omega_{-})] 
v_{s_{3}}(p_{3}) \,,
\label{H_2e2mu_amps}
\end{eqnarray}
%
%
%
where $u_{s}(p)$ and $v_{s}(p)$ are
momentum-dependent Dirac spinors for a fermion
and anti-fermion, respectively.\\

For the decay
\begin{eqnarray}
H(P) \to
\mu^{+}(p_{1}) \,\mu^{-}(p_{2})\, 
\mu^{+}(p_{3}) \,\mu^{-}(p_{4})
\label{H_4mu}
\end{eqnarray}
via the intermediate $Z$ bosons, the amplitude reads
\begin{eqnarray}
{\cal M}_{H \to \mu^{+}\mu^{-}\mu^{+}\mu^{-}} = 
{\cal M}_{H \to \mu^{+}\mu^{-}\mu^{+}\mu^{-}}^{(1)} - 
{\cal M}_{H \to \mu^{+}\mu^{-}\mu^{+}\mu^{-}}^{(2)}
\end{eqnarray}
with ${\cal M}_{H \to \mu^{+}\mu^{-}\mu^{+}\mu^{-}}^{(1)}$ as (\ref{H_2e2mu_amps}) but with
$m_{e} \to m_{\mu}$ in the spinor wave functions
and
\begin{eqnarray}
{\cal M}^{(2)}_{H \to \mu^{+}\mu^{-}\mu^{+}\mu^{-}} 
= \left. {\cal M}^{(1)}_{H \to \mu^{+}\mu^{-}\mu^{+}\mu^{-}} \right|
_{p_{1}, s_{1} \leftrightarrow p_{3}, s_{3}}\,.
\label{H_4mu_amps}
\end{eqnarray}

Above
\begin{eqnarray}
\omega_{\pm} = 
\frac{1}{2} \left(\mathbb{1}_{4 \times 4} \pm \gamma_5 \right)\,, 
\quad
\mathbb{1}_{4 \times 4} = 
\left( \begin{array}{cc}
\mathbb{1} & \;\, 0 \\ 
0 & \;\, \mathbb{1} \\
\end{array} \right)\,,
\quad
\mathbb{1} = 
\left( \begin{array}{cc}
1 & \;\, 0 \\ 
0 & \;\, 1 \\
\end{array} \right)\,,
\end{eqnarray}
$\omega_{\pm}$ are the right- and left-handed chirality projectors,
$\mathbb{1}$ is the identity matrix.
In the 
Dirac-Pauli representation, the gamma matrices 
$\gamma^{\mu}$ ($\mu = 0, 1, 2, 3$) and $\gamma^{5}$ read
\begin{eqnarray}
&&
\gamma^{0} = \gamma_{0} = 
\left( \begin{array}{cc}
\mathbb{1} & \;\, 0 \\ 
0 & \;\, -\mathbb{1} \\
\end{array} \right)\,, \quad
\gamma^{k} = -\gamma_{k} = 
\left( \begin{array}{cc}
0 & \;\, \sigma_{k} \\ 
-\sigma_{k} & \;\, 0 \\
\end{array} \right)\,,
\label{gamma_matrices}\\
&&
\gamma_{5} = \gamma^{5} = i \gamma^{0} \gamma^{1} \gamma^{2} \gamma^{3} = 
\left( \begin{array}{cc}
0 & \;\, \mathbb{1} \\ 
\mathbb{1} & \;\, 0 \\
\end{array} \right)\,.
\label{gamma5}
\end{eqnarray}
In (\ref{gamma_matrices}) $k$ runs from 1 to 3 and the $\sigma_{k}$ are Pauli matrices.

In the Born approximation the relevant coupling constants read [see Appendices~A of \cite{Denner:1991kt,Denner:1994xt}]:
\begin{eqnarray}
&&g_{HZZ} = \frac{M_W}{c^2_{\rm W} s_{\rm W}} =\frac{M_Z}{c_{\rm W} s_{\rm W}}  \,,\\
&&g_{Z\ell\ell}^{+} = -\frac{s_{\rm W}}{c_{\rm W}} Q_\ell \,,\\
&&g_{Z\ell\ell}^{-} = -\frac{s_{\rm W}}{c_{\rm W}} Q_\ell + \frac{I_{{\rm W},\ell}^3}{s_{\rm W} c_{\rm W}}\,.
\end{eqnarray}
Above $c_{\rm W}^2 = 1 - s_{\rm W}^2 = M_W^2/M_Z^2$, 
$Q_\ell = -1$, 
and $I_{{\rm W},\ell}^3 = -1/2$.


The usual Dirac spinors for the fermion and anti-fermion
with momentum $p$ and spin 
in the $\pm z$ direction for $s = \pm 1/2$ 
(spin ``up'' or ``down'') are
\begin{eqnarray}
&&u_{s}(p) =
\sqrt{E + m_{\ell}}
\left( \begin{array}{c}
\chi_{s}^{(1)}
\vspace{0.2cm}\\
\dfrac{\bsigma \cdot \bp}{p^{0} + m_{\ell}}\, \chi_{s}^{(1)} \\
\end{array} \right)\,, 
\label{spinor_usual_u}\\
&&v_{s}(p) =
\sqrt{E + m_{\ell}}
\left( \begin{array}{c}
\dfrac{\bsigma \cdot \bp}{p^{0} + m_{\ell}}\, \chi_{s}^{(2)}
\vspace{0.2cm}\\
\chi_{s}^{(2)}\\
\end{array} \right)\,,
\label{spinor_usual_v}
\end{eqnarray}
where $E = \sqrt{|\bp|^{2} + m_{\ell}^{2}}$,
\begin{equation}
\begin{split}
\bsigma \cdot \bp = \sigma_{k} \, p_{k} =
\left( \begin{array}{cc}
p_{3} & \;\, p_{1}-i p_{2}\\ 
p_{1}+i p_{2} & \;\, -p_{3} \\
\end{array} \right)\,.
\end{split}
\label{sig_dot_p}
\end{equation}
Here the two-spinors $\chi$ are
\begin{equation}
\chi_{1/2}^{(1)} = 
\left( \begin{array}{c}
1 \\
0 
\end{array} \right)\,, \quad
\chi_{-1/2}^{(1)} = 
\left( \begin{array}{c}
0 \\
1 
\end{array} \right)\,,
\label{chi_s1}
\end{equation}
where $\chi_{1/2}$ corresponds to a ``spin up'' state
and $\chi_{-1/2}$ to a ``spin down'' state.
In (\ref{spinor_usual_v}) we have
\begin{equation}
\chi_{s}^{(2)} = \chi_{-s}^{(1)}\,.
\label{chi_s2_aux}
\end{equation}
The helicity spinors can be obtained as follows:
%
\begin{eqnarray}
u_{s'}^{(h)}(p) = 
B^{(a)}(\bhp) \,u_{s}(p) \,,
\end{eqnarray}
where $B^{(a)}(\bhp) = ( B^{(a)}_{s's}(\bhp) )$ is defined by (A13) of \cite{Klusek-Gawenda:2017lgt}
\begin{equation}
\begin{split}
B^{(a)}(\bhp) = \left(B^{(a)}_{s's}(\bhp)\right) =
\left( \begin{array}{cc}
\cos\frac{\theta}{2} & \;\, -\sin\frac{\theta}{2}\; e^{-i \phi}\\ 
\sin\frac{\theta}{2}\; e^{i \phi} & \;\, \cos\frac{\theta}{2} \\
\end{array} \right).
\end{split}
\label{B_07}
\end{equation}

Then, we have for the $u$-type spinor
\begin{eqnarray}
&&u_{ 1/2}^{(h)}(p) =  \cos\frac{\theta}{2}               \,u_{ 1/2}(p) 
                    +  \sin\frac{\theta}{2} \,e^{i\phi}   \,u_{-1/2}(p) \,, \nonumber \\
&&u_{-1/2}^{(h)}(p) = -\sin\frac{\theta}{2} \,e^{-i\phi}  \,u_{ 1/2}(p) 
                    +  \cos\frac{\theta}{2}               \,u_{-1/2}(p) \,,
\label{spinor_usual_hel_u}
\end{eqnarray}
and for the $v$-type spinor
%
\begin{eqnarray}
&&v_{ 1/2}^{(h)}(p) = -\sin\frac{\theta}{2} \,e^{-i\phi}  \,v_{ 1/2}(p) 
                    +  \cos\frac{\theta}{2}               \,v_{-1/2}(p) \,, \nonumber \\
&&v_{-1/2}^{(h)}(p) =  \cos\frac{\theta}{2}               \,v_{ 1/2}(p) 
                    +  \sin\frac{\theta}{2} \,e^{i\phi}   \,v_{-1/2}(p) \,.
\label{spinor_usual_hel_v}
\end{eqnarray}
Here, we 
assumed that the helicity states for 
fermion and anti-fermion
are both taken of the same type, e.g., of type ($a$); 
see Appendix A of \cite{Klusek-Gawenda:2017lgt}.

For the adjoint spinor we have
\begin{eqnarray}
\bar{u}_{s}^{(h)}(p) = u_{s}^{(h)\dagger}(p)\, \gamma^{0}
               = [(u_{s}^{(h)}(p))^{\ast}]^{T}\, \gamma^{0} \,.
\label{spinor_ubar}
\end{eqnarray}
The normalization of the orthogonal 
four-spinors $u$ and $v$ is
\begin{eqnarray}
&&\bar{u}_{s}(p)\, u_{s'}(p) = 2m_{\ell} \,\delta_{ss'}\,,\nonumber \\
&&\bar{v}_{s}(p)\, v_{s'}(p) = -2m_{\ell} \,\delta_{ss'}\,,
\end{eqnarray}
where $s, s' \in \{+1/2,-1/2 \}$.
The completeness relations (or polarization sum rules) are
\begin{eqnarray}
&&\sum_{s} u_{s}(p)\, \bar{u}_{s}(p) = \slash{p} + m_{\ell}\,,\nonumber \\
&&\sum_{s} v_{s}(p)\, \bar{v}_{s}(p) = \slash{p} - m_{\ell}\,,
\label{B_15}
\end{eqnarray}
where $\slash{p} = \gamma^{\mu} p_{\mu}$.

\subsubsection{Amplitude squared from \textsc{FeynCalc}}
\label{subsec_Hto4L_FeynCalc}

Using \textsc{FeynCalc} \cite{FeynCalc,Mertig:1990an,Shtabovenko:2020gxv} 
we obtain the matrix element squared
for the decay $H \to Z^*Z^* \to e^{+}e^{-}\mu^{+}\mu^{-}$ 
(\ref{H_2e2mu}) as follows:
\begin{eqnarray}
|{\cal M}_{H \to e^{+}e^{-}\mu^{+}\mu^{-}}|^{2} &=&
{\cal M}_{H \to e^{+}e^{-}\mu^{+}\mu^{-}}
({\cal M}_{H \to e^{+}e^{-}\mu^{+}\mu^{-}})^{\ast} \nonumber \\ 
&=& \left. (e^{3}\,g_{HZZ})^{2} \,
\frac{1}{((p_1 + p_2)^2 - M_Z^2)^{2} + M_Z^{2} \Gamma_Z^{2}}\;
\frac{1}{((p_3 + p_4)^2 - M_Z^2)^{2} + M_Z^{2} \Gamma_Z^{2}}
             \right. \nonumber \\ 
&& \left. \times 16 \Bigl[ 4 \,(g_{Z\ell\ell}^{+})^{2} (g_{Z\ell\ell}^{-})^{2}\, 
      m_{e}^{2} \, m_{\mu}^{2} \right. \nonumber \\ 
&& \left. + \,g_{Z\ell\ell}^{+} \,g_{Z\ell\ell}^{-} 
      \left( (g_{Z\ell\ell}^{+})^{2}+(g_{Z\ell\ell}^{-})^{2} \right)
      m_{\mu}^{2} \; p_{1} \cdot p_{2}  \right. \nonumber \\ 
&& \left. + \,g_{Z\ell\ell}^{+} \,g_{Z\ell\ell}^{-} 
      \left( (g_{Z\ell\ell}^{+})^{2}+(g_{Z\ell\ell}^{-})^{2} \right) 
      m_{e}^{2} \; p_{3} \cdot p_{4} \right. \nonumber \\ 
&& \left. + \,2 \,(g_{Z\ell\ell}^{+})^{2} (g_{Z\ell\ell}^{-})^{2} \,
      p_{1} \cdot p_{4} \,p_{2} \cdot p_{3} \right. \nonumber \\ 
&& \left. + \left( (g_{Z\ell\ell}^{+})^{4}+(g_{Z\ell\ell}^{-})^{4} \right) \,
      p_{1} \cdot p_{3} \,p_{2} \cdot p_{4} \Bigr] \right. .
\label{FC_2e2mu}
\end{eqnarray}

The matrix element squared for the process
$H \to Z^*Z^* \to \mu^{+}\mu^{-}\mu^{+}\mu^{-}$ 
(\ref{H_4mu}) can be written as
(including identity factor 1/4)
\begin{eqnarray}
\frac{1}{4} |{\cal M}_{H \to \mu^{+}\mu^{-}\mu^{+}\mu^{-}}|^{2} =
\frac{1}{4} \left( |{\cal M}_{11}|^{2} + |{\cal M}_{22}|^{2} + 
|{\cal M}_{{\rm int}}|^{2} \right)\,,
\label{FC_4mu}
\end{eqnarray}
where
\begin{eqnarray}
|{\cal M}_{11}|^{2} &=&
{\cal M}^{(1)}
({\cal M}^{(1)})^{\ast} \nonumber \\ 
&=& \left. (e^{3}\,g_{HZZ})^{2} \,
\frac{1}{((p_1 + p_2)^2 - M_Z^2)^{2} + M_Z^{2} \Gamma_Z^{2}}\;
\frac{1}{((p_3 + p_4)^2 - M_Z^2)^{2} + M_Z^{2} \Gamma_Z^{2}}
             \right. \nonumber \\ 
&& \left. \times 16 \Bigl[ 4 \,(g_{Z\ell\ell}^{+})^{2} (g_{Z\ell\ell}^{-})^{2}\, 
      m_{\mu}^{2} \, m_{\mu}^{2} \right. \nonumber \\ 
&& \left. + \,g_{Z\ell\ell}^{+} \,g_{Z\ell\ell}^{-} 
      \left( (g_{Z\ell\ell}^{+})^{2}+(g_{Z\ell\ell}^{-})^{2} \right)
      m_{\mu}^{2} \; p_{1} \cdot p_{2}  \right. \nonumber \\ 
&& \left. + \,g_{Z\ell\ell}^{+} \,g_{Z\ell\ell}^{-} 
      \left( (g_{Z\ell\ell}^{+})^{2}+(g_{Z\ell\ell}^{-})^{2} \right) 
      m_{\mu}^{2} \; p_{3} \cdot p_{4} \right. \nonumber \\ 
&& \left. + \,2 \,(g_{Z\ell\ell}^{+})^{2} (g_{Z\ell\ell}^{-})^{2} \,
      p_{1} \cdot p_{4} \;p_{2} \cdot p_{3} \right. \nonumber \\ 
&& \left. + \left( (g_{Z\ell\ell}^{+})^{4}+(g_{Z\ell\ell}^{-})^{4} \right) \,
      p_{1} \cdot p_{3} \;p_{2} \cdot p_{4} \Bigr] \right. ,
\label{FC_4mu_M11}\\
|{\cal M}_{22}|^{2} &=&
{\cal M}^{(2)}
({\cal M}^{(2)})^{\ast}
= \left. |{\cal M}_{11}|^{2} \right|
_{p_{1} \leftrightarrow p_{3}}\,,
\label{FC_4mu_M22}
\end{eqnarray}

\begin{eqnarray}
|{\cal M}_{\rm int}|^{2} 
&=&
{\cal M}^{(1)}
({\cal M}^{(2)})^{\ast} + {\cal M}^{(2)}
({\cal M}^{(1)})^{\ast}\nonumber\\
&=& \left. (e^{3}\,g_{HZZ})^{2} \,
             \right. \nonumber \\ 
&& \left. \times 8 \, \Big\lbrace
g_{Z\ell\ell}^{+} \, g_{Z\ell\ell}^{-} \, 
m_{\mu}^{2} \,
\Bigl[ \left( (g_{Z\ell\ell}^{+})^{2}+(g_{Z\ell\ell}^{-})^{2} \right)\,
 \left( p_{1} \cdot p_{4} 
       +p_{1} \cdot p_{2}
       +p_{2} \cdot p_{3}
       +p_{3} \cdot p_{4}
 \right) \right. \nonumber \\ 
&& \left. - 2\, g_{Z\ell\ell}^{+} \, g_{Z\ell\ell}^{-}\, 
\left( p_{2} \cdot p_{4} + p_{1} \cdot p_{3} - 2 m_{\mu}^{2} \right)
\Bigr] 
+ 2 \left( (g_{Z\ell\ell}^{+})^{4}+(g_{Z\ell\ell}^{-})^{4} \right)\,
p_{1} \cdot p_{3}\; p_{2} \cdot p_{4} \Big\rbrace  \right. \nonumber \\
&& \left. \times 
\mathrm{Re} 
\Bigl[ \Bigl( 
\frac{1}{(p_1 + p_2)^2 - M_Z^2 + i M_Z \Gamma_Z}\;
\frac{1}{(p_3 + p_4)^2 - M_Z^2 + i M_Z \Gamma_Z} \right. \nonumber \\
&& \left. \times 
\frac{1}{(p_2 + p_3)^2 - M_Z^2 - i M_Z \Gamma_Z}\;
\frac{1}{(p_1 + p_4)^2 - M_Z^2 - i M_Z \Gamma_Z} \Bigl) \right. \nonumber \\
&& \left. + \Bigl( 
\frac{1}{(p_1 + p_2)^2 - M_Z^2 - i M_Z \Gamma_Z}\;
\frac{1}{(p_3 + p_4)^2 - M_Z^2 - i M_Z \Gamma_Z} \right. \nonumber \\
&& \left. \times 
\frac{1}{(p_2 + p_3)^2 - M_Z^2 + i M_Z \Gamma_Z}\;
\frac{1}{(p_1 + p_4)^2 - M_Z^2 + i M_Z \Gamma_Z} \Bigl) 
\Bigl] \right. .
\label{FC_4mu_Mint} 
\end{eqnarray}
%

\subsubsection{Numerical results and comparison with other MC results}
\label{subsec_Hto4L_NumResults}

In Table~\ref{Tab.4Fermi_TGenDecay} we present the results
of partial decay width (\ref{Gamma_H4L})
using \classMark{TGenFoamDecay} for different generator set-up;
that is different number of events, cells ($nCells$),
and samplings per cell ($nSampl$).
In the calculation we take $M_{H} = 125.1$~GeV.
For the electromagnetic coupling constant
we use 
the expression
derived from the Fermi constant $G_{F}$, the muon decay constant,
according to
$\alpha_{G_{\mu}} = \frac{\sqrt{2} G_{F} M_{W}^{2}}{\pi} 
(1-\frac{M_{W}^{2}}{M_{Z}^{2}})$, 
i.e. $\alpha_{G_{\mu}} \simeq 1/132.184$ 
instead of $\alpha = e^{2}/(4 \pi) \simeq 1/137.036$.
\begin{table}[!h]
\begin{center}
\begin{tabular}{ |c|c|c|c|c| }
\hline
 \# Events & $nCells$ & $nSampl$ & $\Gamma(H \to 2e 2\mu)$ [keV] & $\Gamma(H \to 4\mu)$ [keV] \\ \hline 
 $10^6$ & $10^{3}$ & $10^{3}$ & $0.240418 \pm 0.000103$ & $0.132640 \pm 0.000066$ \\ \hline
 $10^6$ & $10^{4}$ & $10^{3}$ & $0.240193 \pm 0.000085$ & $0.132645 \pm 0.000065$\\ \hline
 $10^6$ & $10^{4}$ & $10^{4}$ & $0.240150 \pm 0.000084$ & $0.132718 \pm 0.000063$\\ \hline
 $10^7$ & $10^{3}$ & $10^{3}$ & $0.240244 \pm 0.000033$ & $0.132686 \pm 0.000021$\\ \hline
 $10^7$ & $10^{4}$ & $10^{3}$ & $0.240224 \pm 0.000027$ & $0.132645 \pm 0.000021$\\ \hline
 $10^7$ & $10^{4}$ & $10^{4}$ & $0.240238 \pm 0.000027$ & $0.132652 \pm 0.000020$ \\ \hline
\end{tabular}
\end{center}
\caption{The values of $\Gamma(H \to 4 \ell)$
obtained by adaptive \classMark{TGenFoamDecay} integration
for different FOAM parameter settings.
Results were obtained with the amplitudes
described in Section~\ref{subsec_Hto4L_FeynCalc}.}
 \label{Tab.4Fermi_TGenDecay}
\end{table}



Comparison with other MC results.

There are \textsc{Hto4l} MC results from \cite{Boselli:2015aha},
$\Gamma(H \to e^+ e^- \mu^+ \mu^-) = 0.24165(2)$~keV and
$\Gamma(H \to \mu^+ \mu^- \mu^+ \mu^-) = 0.13325(2)$~keV,
both including the NLO electroweak corrections; 
see Table~2 of \cite{Boselli:2015aha}.
We can see from the lower plot of Figure~3 in \cite{Boselli:2015aha} 
that the LO (tree-level) $4 \mu$ result\footnote{In the Hto4l LO results,
effectively the ``$\alpha_{G_\mu}$ scheme'' was used 
with $\alpha_{G_{\mu}} = 1/132.4528503$.} 
is about 2\,\% smaller than the NLO result.

The \textsc{Prophecy4F} MC results (LO) for $M_{H} = 140$~GeV 
from Table~1 of \cite{Bredenstein:2006rh} are:
$\Gamma(H \to e^+ e^- \mu^+ \mu^-) = 1.2349(4)$~keV,
$\Gamma(H \to \mu^+ \mu^- \mu^+ \mu^-) = 0.6555(2)$~keV.
We get for $10^{7}$ events, $nCells = 10^{3}$, 
$nSampl = 10^{3}$, and for $M_{H} = 140$~GeV:
$\Gamma(H \to e^+ e^- \mu^+ \mu^-) = 1.2337$~keV,
$\Gamma(H \to \mu^+ \mu^- \mu^+ \mu^-) = 0.6546$~keV.

\begin{figure}[!ht]
\includegraphics[width=8.cm]{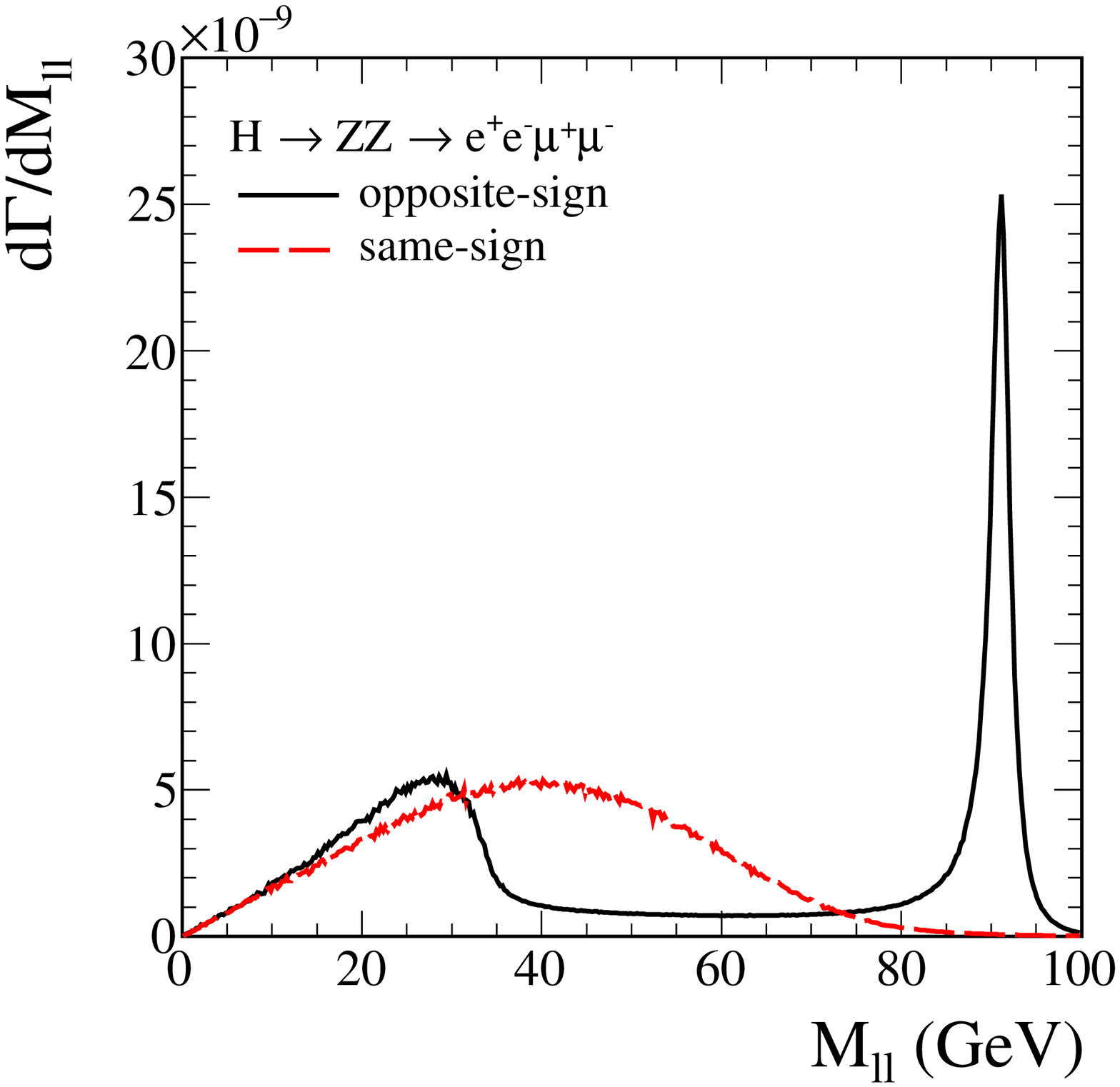}
\includegraphics[width=8.cm]{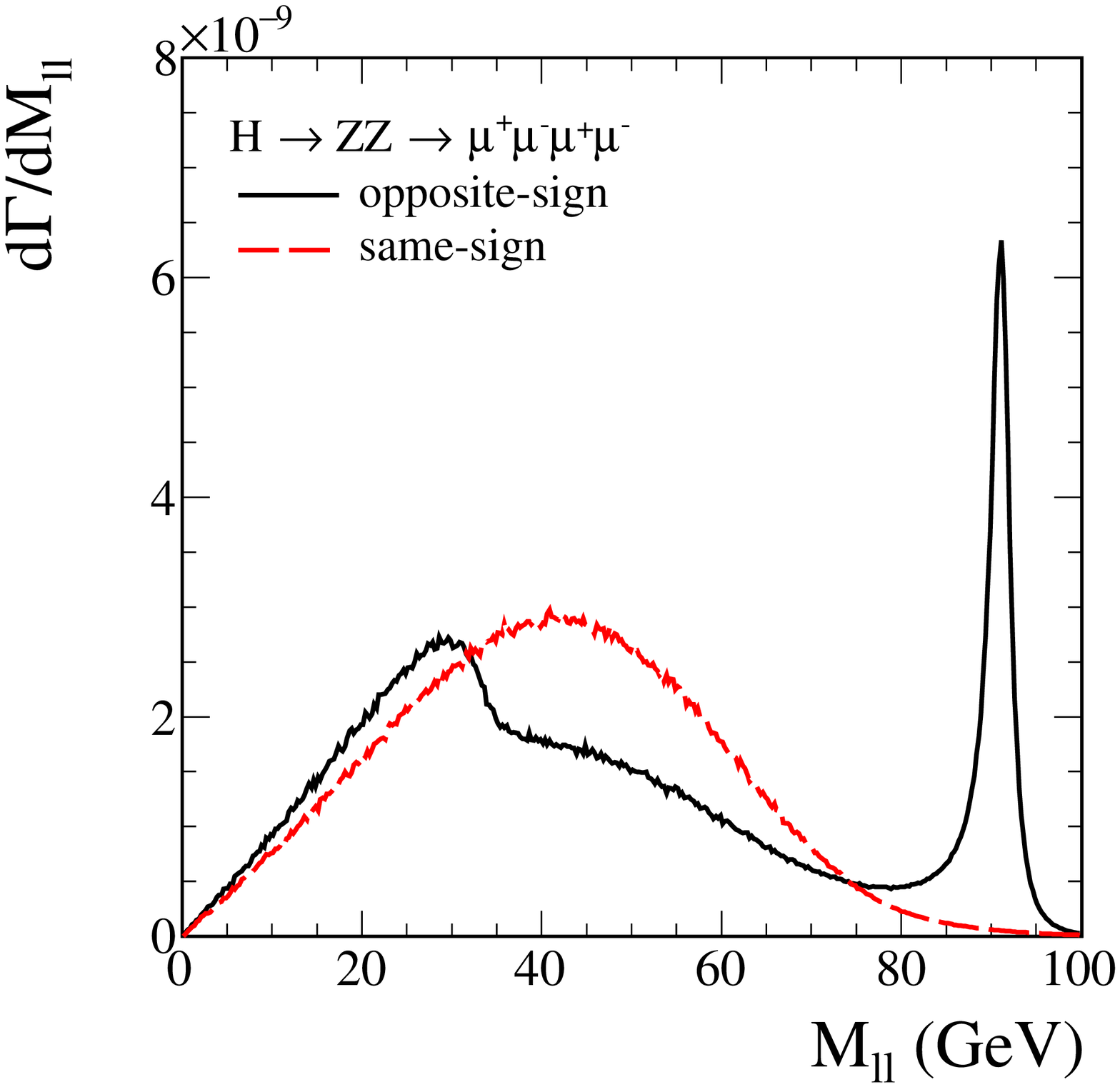}
\caption{The distributions in $\ell \ell$ invariant mass 
for the decay $H \to Z^*Z^* \to e^+ e^- \mu^+ \mu^-$ (left panel)
and $H \to Z^*Z^* \to \mu^+ \mu^- \mu^+ \mu^-$ (right panel).}
\label{fig:mumumu_1}
\end{figure}

\begin{figure}
\includegraphics[width=8.cm]{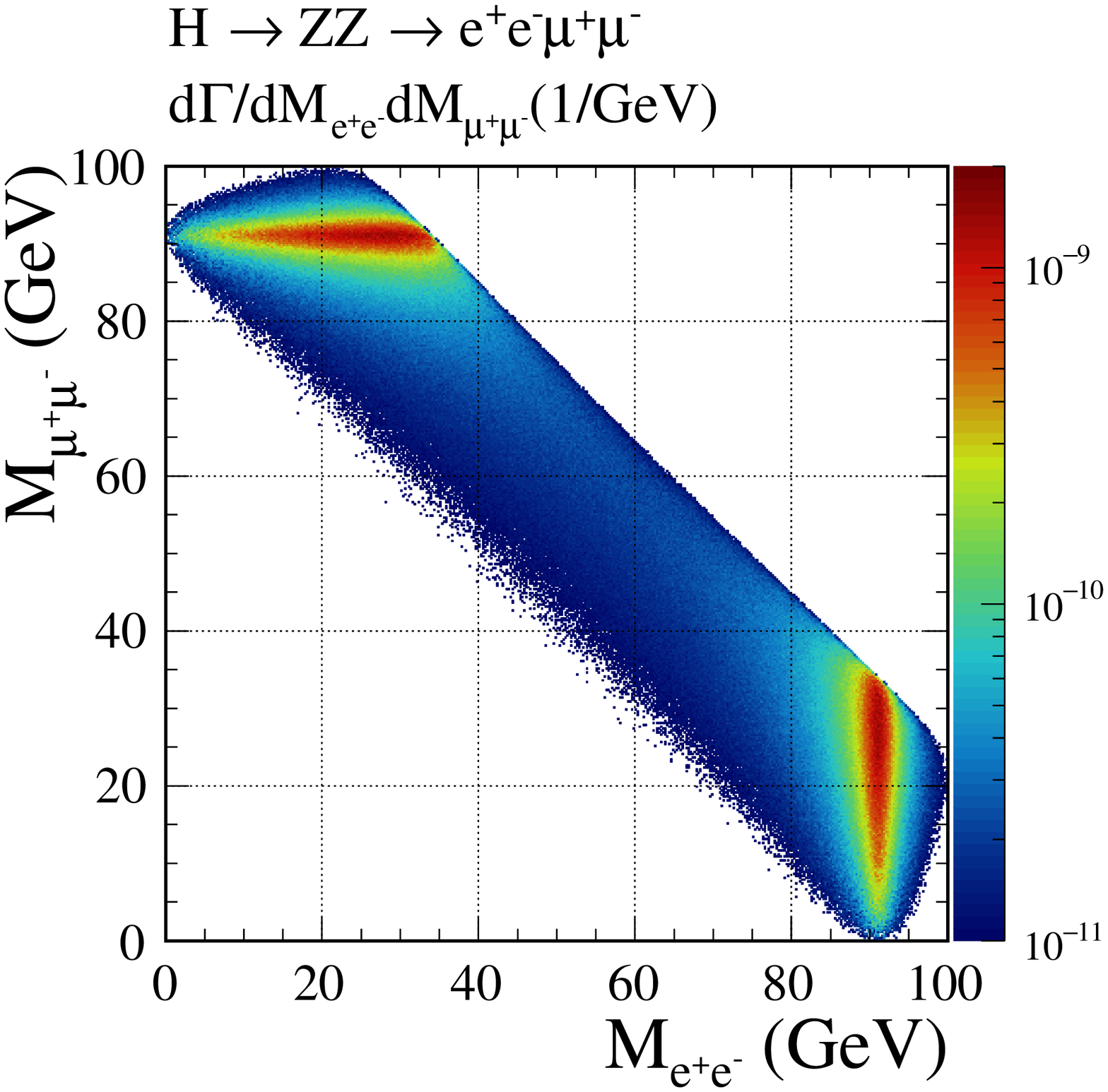}
\includegraphics[width=8.cm]{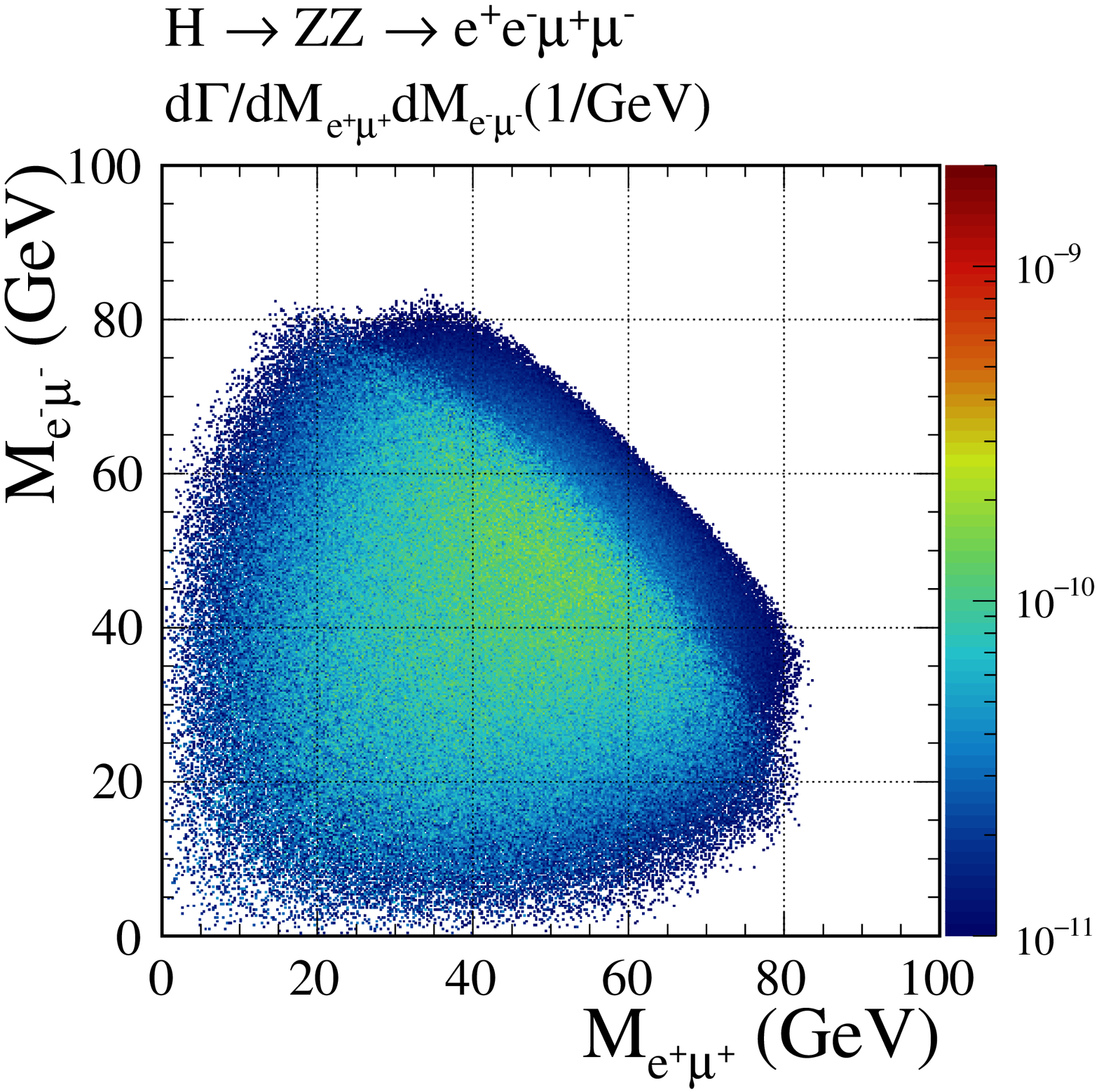}\\
\includegraphics[width=8.cm]{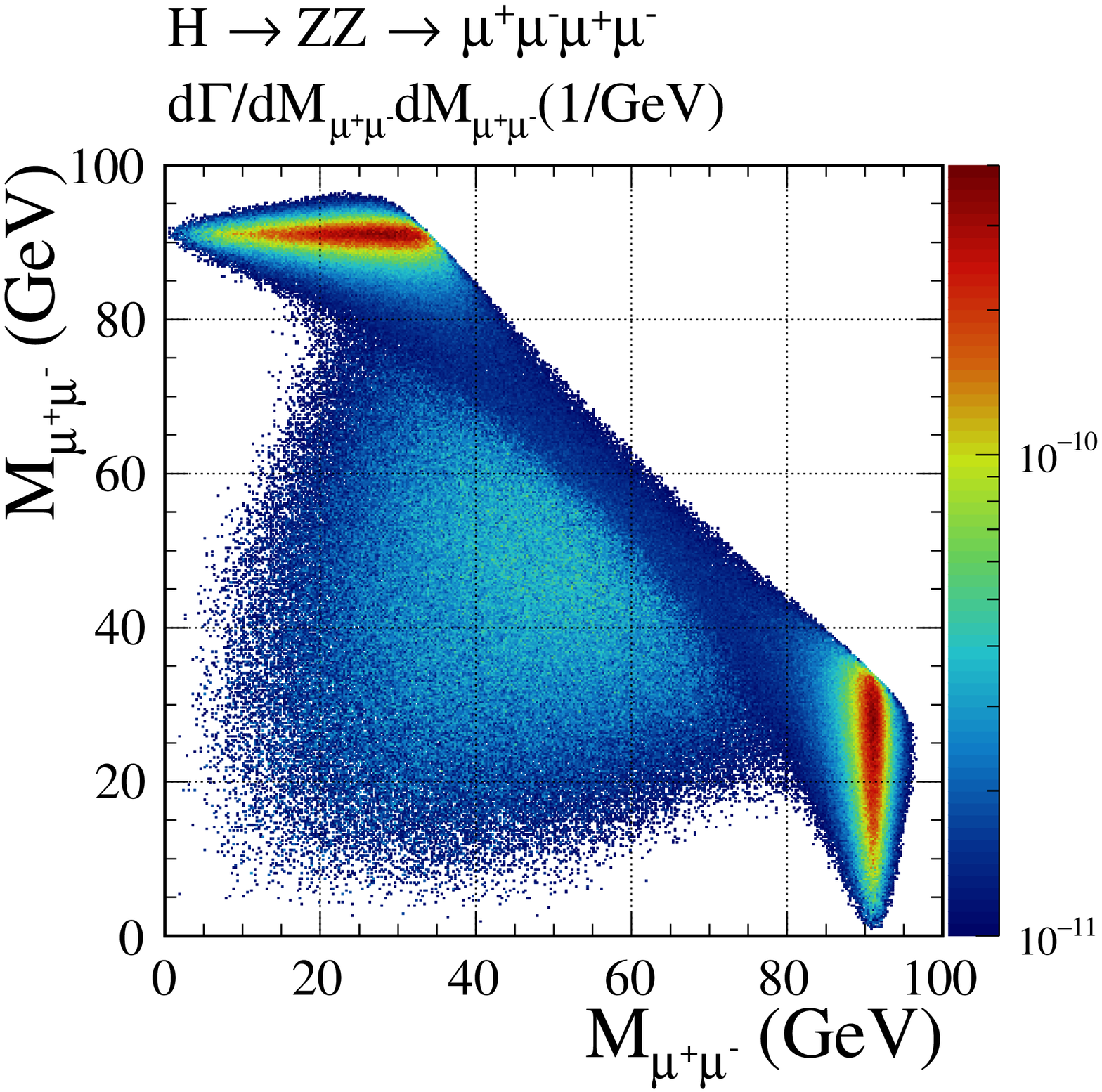}
\includegraphics[width=8.cm]{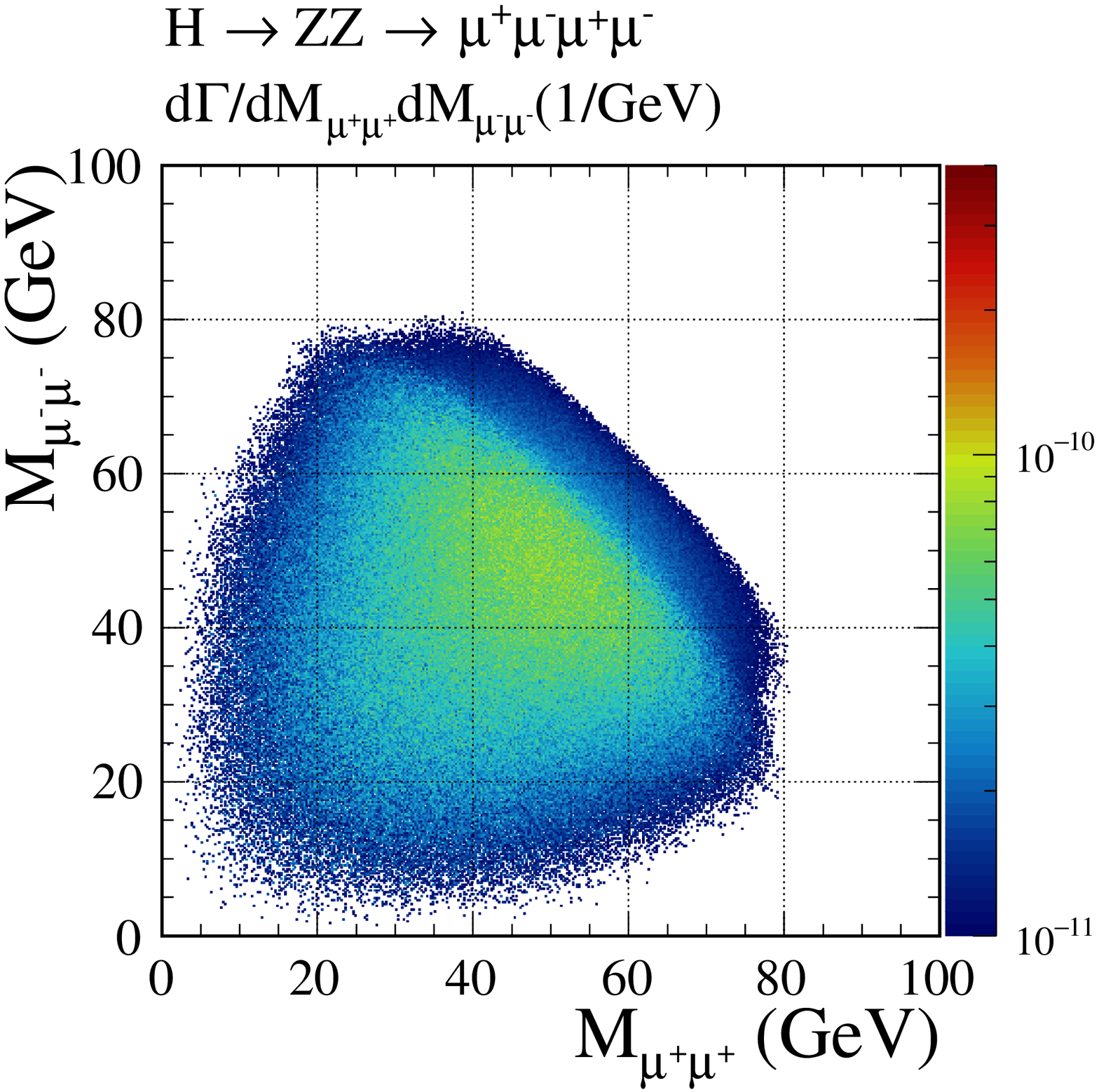}
\caption{Two-dimensional distributions in dilepton invariant masses 
for the decay $H \to Z^*Z^* \to e^+ e^- \mu^+ \mu^-$
(top panels) and $H \to Z^*Z^* \to \mu^+ \mu^- \mu^+ \mu^-$
(bottom panels).}
\label{fig:eemumu_2}
\end{figure}

\begin{figure}[!h]
\includegraphics[width=8.cm]{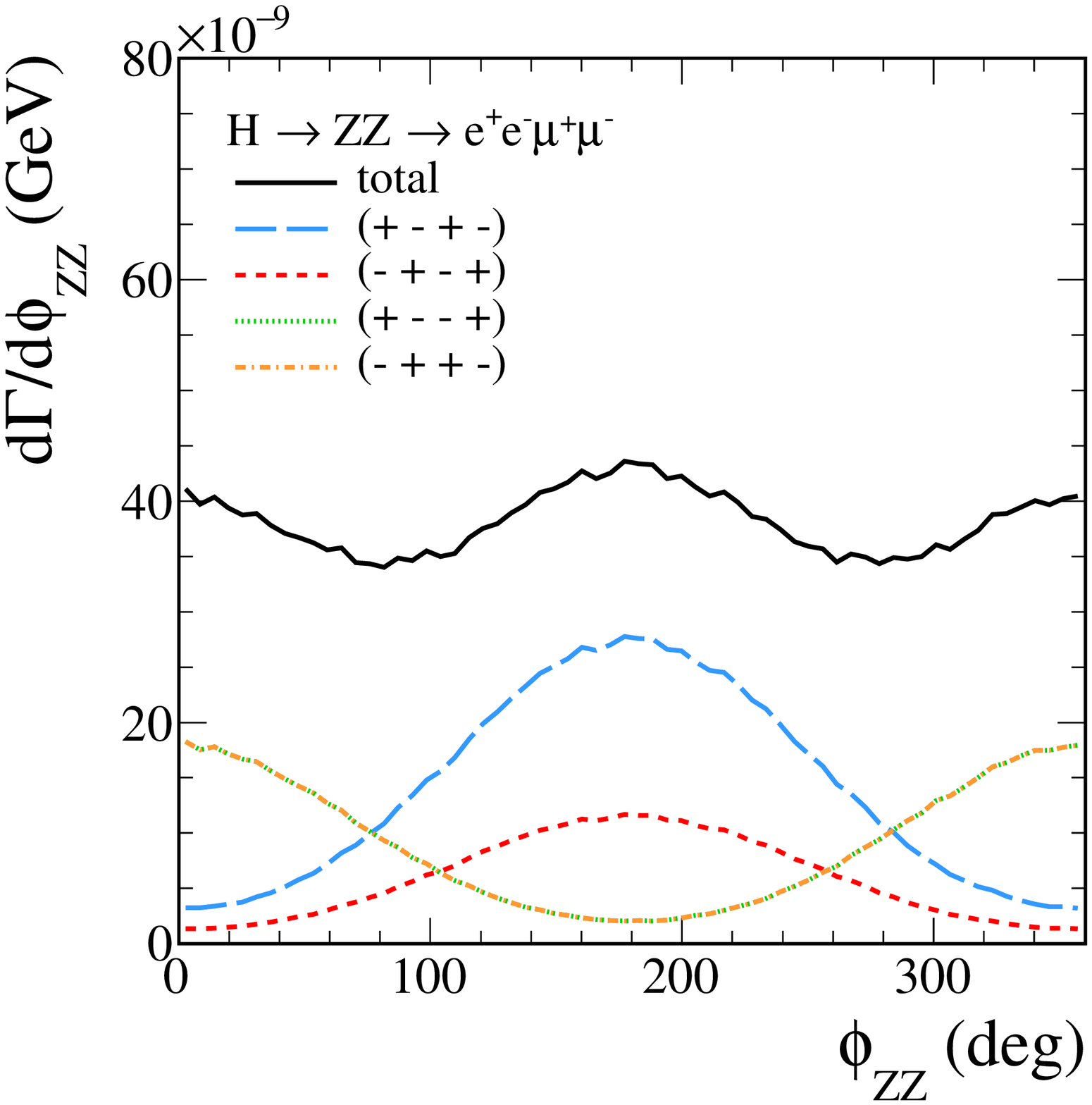}
\caption{
The distribution in $\phi_{ZZ}$ (\ref{phi_ZZ})
for the decay $H \to Z^*Z^* \to e^+ e^- \mu^+ \mu^-$.
Contributions of different helicities
($2 s_{1}$, $2 s_{2}$, $2 s_{3}$, $2 s_{4}$) 
of the outgoing leptons are shown separately.
}
\label{fig:eemumu_aux1}
\end{figure}

Examples of differential distributions
are shown in Figs.~\ref{fig:mumumu_1}--\ref{fig:eemumu_aux1}.
Results are obtained with the method (A) 
(see Section~\ref{subsec_two_modes_of_using}) by generating $10^{7}$ events. 
In the calculation we take $M_{H} = 125.1$~GeV.

In Fig.~\ref{fig:mumumu_1} we show the dilepton invariant mass distributions 
for the $H \to Z^*Z^* \to e^+ e^- \mu^+ \mu^-$ 
and $H \to Z^*Z^* \to \mu^+ \mu^- \mu^+ \mu^-$ decay processes.
Figure~\ref{fig:eemumu_2} shows results 
for the two-dimensional distributions.
We can observe a different pattern of the distributions.
Because of two diagrams in the $H \to 4\mu$ decay
the interference term contributes there.

Interesting observable suitable for the spin-parity assignment 
is $\phi_{ZZ}$, the angle between the decay planes 
of the virtual $Z$ bosons in the $H$ rest frame.
For the $\phi_{ZZ}$ angle we use the definition
\cite{Bredenstein:2006rh,Boselli:2015aha}
(see also \cite{Mikhasenko:2020qor})
\begin{eqnarray}
&&\cos \phi_{ZZ} = 
\frac{(\bpab \times \bpaa) \cdot (\bpbb \times \bpbc)}
{|\bpab \times \bpaa| |\bpab \times \bpbc|}\,, \nonumber \\
&&{\rm sgn}(\sin \phi_{ZZ}) = 
  {\rm sgn}\lbrace \bpab \cdot [(\bpab \times \bpaa) \times (\bpab \times \bpbc)] \rbrace\,,
\label{phi_ZZ}
\end{eqnarray}
where $\bpab = \bpaa + \bpbb$.

In Fig.~\ref{fig:eemumu_aux1} we present 
the distribution in $\phi_{ZZ}$ for $H \to Z^*Z^* \to e^+ e^- \mu^+ \mu^-$.
We label the results for different helicity
terms as ($2 s_{1}$, $2 s_{2}$, $2 s_{3}$, $2 s_{4}$)
where $s_{i} = \pm \frac{1}{2}$.
The distributions in $\phi_{\ell}$ and $\cos\theta_{\ell}$ 
are constants for both the decay channels.

We have checked that the helicity amplitudes of $H \to 2e 2\mu$
with the helicity spinors
(\ref{spinor_usual_hel_u}) and (\ref{spinor_usual_hel_v})
in the massless limit are consistent with those given by 
Eqs.~(9)--(13), (D4), (D5) in \cite{Campbell:2013una}
and by Eq.~(A1) in~\cite{He:2019kgh}
that uses massless helicity spinor formalism 
(see also Eqs.~(14) and (15) of \cite{Dixon:1996wi}).

\section{Conclusions}

We have presented a new Monte Carlo library \textsc{Decay} 
and examples of generators that can generate events 
for the decay of any particle into higher multiplicity 
final state and provide an easy way of adaptive integration 
over Lorentz Invariant Phase Space. 
They integrate with common in High Energy Physics software - ROOT 
and have a compatible interface with other ROOT generators. 

We have also presented a few examples of applications of these tools,
namely, decay into two particles, decay into three
particles within the $4$-Fermi theory of $\mu$ decay, and 
the Standard Model Higgs boson decay into four leptons 
in leading order. Special attention has been devoted 
to the Higgs boson decay, where we have presented 
some interesting results 
which could be studied experimentally.
All the tests have turned out to be compatible
with results known from the literature.

We hope that in the near future our generator 
will be a part of the commonly available ROOT software.

\acknowledgments
This work was partially supported by the Polish National Science Centre 
under Grant No. 2018/31/B/ST2/03537. 
The work of R.K. was supported by the GACR grant GA19-06357S 
and Masaryk University grant MUNI/A/0885/2019. 
R.K. also thank the SyMat COST Action (CA18223) for partial support.

The authors are grateful to Carlo Carloni Calame and Fulvio Piccinini
for a help in comparing our results with the results of their original study.

\bibliography{refs}

\end{document}